\documentclass[sigconf]{acmart}
\AtBeginDocument{%
  \providecommand\BibTeX{{%
    \normalfont B\kern-0.5em{\scshape i\kern-0.25em b}\kern-0.8em\TeX}}}

\setcopyright{acmcopyright}
\copyrightyear{2020}
\acmYear{2020}
\acmDOI{10.1145/1122445.1122456}

\acmConference[KDD - MLF - 2020]{KDD Workshop on Machine Learning in Finance}{August 24, 2020}{San Diego, CA}
\acmBooktitle{KDD Workshop on Machine Learning in Finance, August 24, 2020, San Diego, CA}
\usepackage{times}
\usepackage{soul}
\usepackage{url}
\usepackage[utf8]{inputenc}
\usepackage{graphicx}
\usepackage{amsmath}
\usepackage{amsthm}
\usepackage{booktabs}
\usepackage{float}
\usepackage{algorithmic}
\usepackage[ruled]{algorithm2e}
\usepackage{array}

\begin{document}

\title{Explainable Clustering and Application to Wealth Management Compliance}

\author{Enguerrand Horel}
\email{ehorel@stanford.edu}
\author{Kay Giesecke}
\email{giesecke@stanford.edu}
\affiliation{%
  \institution{Stanford University}
}

\author{Victor Storchan}
\email{victor.storchan@jpmchase.com}
\author{Naren Chittar}
\email{naren.chittar@jpmchase.com}
\affiliation{%
 \institution{J.P. Morgan}
}


\begin{abstract}
Many applications from the financial industry successfully leverage clustering algorithms to reveal meaningful patterns among a vast amount of unstructured financial data. However, these algorithms suffer from a lack of interpretability that is required both at a business and regulatory level. In order to overcome this issue, we propose a novel two-steps method to explain clusters. A classifier is first trained to predict the clusters labels, then the Single Feature Introduction Test (SFIT) method is run on the model to identify the statistically significant features that characterize each cluster. We describe a real wealth management compliance use-case that highlights the necessity of such an interpretable clustering method. We illustrate the performance of the method using simulated data and through an experiment on financial ratios of U.S. companies.
\end{abstract}



\maketitle

\section{Introduction and Related Work}

The large amount of data and diversity of sources (financial time series, customers characteristics, financial statements, etc.) from the financial industry, create many applications where unsupervised learning techniques can be leveraged successfully. Clustering methods especially, can find meaningful patterns among this unstructured data and provide support for decision making. Applications of clustering techniques to financial cases can be found in risk analysis \cite{kou2014evaluation}, credit scoring \cite{xiao2016ensemble}, financial time series analysis \cite{gavrilov2000mining}, \cite{dias2015clustering}, portfolio management \cite{lemieux2014clustering} and financial statements' anomalies detection \cite{omanovic2009line}. In all these applications, being able to interpret the obtained clusters and explain the rational behind their construction is necessary to trust them and provide transparency to regulators.

Some previous works tackle the problem of interpreting clusters by visualizing them across two or three dimensions typically found by a PCA analysis \cite{rao1964use}. This has the disadvantage of restricting the number of dimensions used to explain and in addition, the principal components are no more directly interpretable. Another group of methods uses the centroid or a selected set of points to represent the cluster \cite{radev2004centroid}. These methods, while successful in some cases, are very sensitive to the geometry of the clusters. Distinct from these previous methods that directly interpret the clusters, two-steps methods explain the clusters through an interpretable model that learns how to classify them. The cluster assignment of each point can be used to label the data and train a classifier on them. Classification trees \cite{breiman2017classification} are often used in practice \cite{hancock2003supervised}. Because the model has to be interpretable, this prevent the use of a larger class of models such as deep neural networks that could potentially provide better classification accuracy.  Another work proposes to directly generate interpretable tree-based clustering models \cite{bertsimas2018interpretable}. It has the inconvenient of restricting the type of clustering algorithm that can be used and it has been showed that such tree-based algorithm can under-perform other algorithms like K-means in several cases \cite{bertsimas2018interpretable}.

To overcome these limitations, we propose to interpret a classifier trained on the clusters by using the SFIT method \cite{horel2019computationally}. This method identifies the statistically significant features of the model as well as feature interactions of any order in a hierarchical manner. It can be applied to any classification model and is computationally efficient. Hence, by combining a two-steps approach with this general model interpretability method, we do not have to restrict the choice of clustering technique nor the choice of classifier to predict the clusters assignment. This provides a general interpretability framework that can be applied to any clustering algorithm and type of data.

The structure of this paper is as follow. In section 2, we describe a real use-case from the financial industry that illustrates the problem of interpreting clusters and present our approach to solve this problem. We describe in section 3, our approach's key component: the SFIT method used to interpret the cluster classifier. Simulation results confirms the efficiency of our method in section 4. Because the data of the described use-case are highly sensitive, we could not use them directly. As a replacement, we illustrate our method on a dataset of U.S. companies clustered using their financial ratios in section 5.

\section{Explainable clustering for compliance monitoring}
\subsection{An overview of the business case}

%provide outliers profiling to Wealth management compliance team
In banks, Wealth Management teams help customers to meet their financial goals by managing their financial assets. An account is associated to each client and an investment strategy is designed by the account manager according to the risk aversion of the client. Using financial performance metrics as features, the efficiency of an account's strategy can be monitored. Comparison with benchmarks or indices are examples of such performance metrics. 

When an account is under-performing based on these features, it is directed to the Compliance team. As a matter of fact, the Compliance team has to spend a considerable amount of time manually reviewing accounts across all monitoring activities and across all global regions to catch the problematic ones. They typically use criteria learned from previous cases to identify the accounts that will require further investigation. This is usually done in an ad-hoc manner based on a small number of features. As a consequence, systematic and generalizable explanations of why an account was flagged as problematic are missing in most cases.

In order to automatise this procedure and identify all the different cases that result in an under-performing behavior, a clustering algorithm is implemented as a solution. Clustering has the advantage of being performed using a large number of different risk factors. This enables to group accounts that behave similarly and bring understanding on the underlying reasons of poor performance. Clustering accounts can also be used to sample from the cluster distribution and estimate the frequency of one particular type of under-performance. However, in order for this clustering method to be confidently deployed, it has to be fully understandable by the Wealth Management Compliance team. They need to explain what are the significant features that characterize each cluster and ensure that it is aligned with their expertise and domain knowledge.
% They aim at recommending representative underperforming accounts which need compliance attention and escalation while providing explanations that justify their choices.

\subsection{A proposal to mitigate the compliance monitoring challenge}

As mentioned in the previous section, our goal is to design an interpretable clustering capability for Compliance Monitoring teams that can then be used to identify under-performing accounts. Our proposed approach relies on two successive steps. 

\textbf{The Clustering Step}. A clustering algorithm (such as K-means, DBSCAN, or agglomerative clustering, etc.) is run over the set of all under-performing accounts.

\textbf{The Explaining Step}. A label is assigned to every cluster which allows to label the whole set of accounts. We can then train a classifier in a supervised way using this dataset. This classifier learns to predict the cluster of a given account. The SFIT procedure can now be applied on the trained classifier. A single model has been trained to classify all clusters, but we ultimately want to interpret each cluster independently. To interpret a specific cluster, the SFIT method is run using only the accounts belonging to this cluster. This allows us to provide for each cluster, a set of features that are significantly characterizing it.

\section{Presentation of the SFIT method}

SFIT \cite{horel2019computationally}, is a method that assesses the statistical significance and importance of features of machine learning models. It is based on a novel application of a forward-selection approach. Given a trained model and one of its features, it compares the predictive performance of the model that uses only the intercept with the model that uses both the intercept and the feature. The performance difference captures the intrinsic contribution of the feature in isolation which leads to an informative notion of feature importance. This approach has the advantage of being robust to correlation among features. Other advantages of our method include: (1) it does not assume any assumptions on the distribution of the data nor assumptions on the specification of the model; (2) it can be applied to both continuous and categorical types of features; (3) in addition to assessing the contribution of individual features, it can also identify higher order interactions among features in a hierarchical manner.

Formally, a set of $n$ i.i.d. accounts $Z_1,...,Z_n \sim P$ with $Z_i = (X_i, Y_i)$. $X_i$ is a vector of size $(p+1)$ that contains the $p$ features measuring the performance of the account plus an intercept at the first coordinate. The features can be a mix of continuous and categorical variable with the latter assumed to be binary variables through one-hot encoding. $Y_i$ represents the index of the cluster $\{1,2,...,C\}$ where $C$ represents the total number of clusters. 
We randomly split the accounts $\{1,...,n\}$ into two subsets $\mathcal{I}_1$ and $\mathcal{I}_2$ and denote the two corresponding split of the data as $\mathcal{D}_k = \{(X_i,Y_i): i \in \mathcal{I}_k\}$, $k = 1,2$.

We denote by $\hat{\mu}$, the cluster classifier trained on $\mathcal{D}_1$. To evaluate the contribution of the individual feature $j$ removed from the potential interaction that it could have with the remaining features, we define the transformed input vector $X_i^{(1,j)}$ which is obtained from $X_i$ where all entries except for the $j^{th}$ coordinate and the intercept are replaced with $0$. Similarly, $X_i^{(1)}$ is the transformed input vector where all entries except for the intercept are set to zero. This transformed input prevents us from having to refit a new model for each input. Then, given the loss function $L(Y,\mu(X))$ used to train the classifier (like the cross entropy loss for instance), we can define $$\Delta_j^i = \Delta_j(X_i,Y_i)  = (1-\beta)L\Big(Y_{i},\hat{\mu}\big(X_i^{(1)}\big)\Big) - L\Big(Y_{i},\hat{\mu}\big(X_i^{(1,j)}\big)\Big)$$ the difference between $(1-\beta)\%$ of the prediction loss from the model using the intercept term only and the loss from the model using the intercept plus the feature $j$. $(1-\beta)$ times the baseline loss and not the loss value itself is considered to make this test more robust to non-informative variables and control its type-I error. More details about this parameter and how to optimally select can be found in \cite{horel2019computationally}. 

Let's now define $\hat{m}_j$, the metric that is used to assess the significance of feature $j$:
$$\hat{m}_j = \text{median}_{i \in \mathcal{D}_2} \Delta_j^i $$ for $2 \leq j \leq p + 1$.
$\hat{m}_j$ is defined as the median over the inference set $\mathcal{D}_2$ of the differences of predictive performance. Intuitively, $\hat{m}_j$ represents the predictive power of variable $j$ compared to a baseline model that only has an intercept term, the bigger it is, the more predictive power the variable contains. 

The statistic $\hat{m}_j$ is used to construct a finite-sample significance test for feature $j$. We assume that the distribution of $\Delta_j(X,Y)$ conditional on the training set $\mathcal{D}_1$ is continuous and consider its median, 
$$m_j = \text{median}_{(X,Y)}\big[ \Delta_j(X,Y) | \mathcal{D}_1\big].$$
We obtain a finite-sample confidence interval for $m_j$ and perform the following one-sided hypothesis test of significance:
\begin{equation}
\label{eq:test1}
H_0 : m_j \leq 0 \; \text{versus} \; H_1: m_j > 0
\end{equation}
using the statistic $\hat{m}_j$. The validity of this test and its uniformly most powerful property are derived in \cite{horel2019computationally}.

The SFIT procedure is described in details in Algorithm \ref{algo 1}.

\begin{algorithm}
\SetAlgoLined
\KwIn{Model $\mu$, dataset $\mathcal{D}_2 = \{X_i,Y_i\}_{i \in \mathcal{I}_2}$, significance level $\alpha$, $\beta$, function $\textsc{BinomTest}$($\#$successes, $\#$trials, hypothesized probability of success, alternative hypothesis) that outputs the p-value of a binomial test}
\KwOut{Set of first order significant variables $\mathcal{S}_1$ and their related confidence intervals $\mathcal{C}_1$}
$\mathcal{S}_1 = \emptyset$, $\mathcal{C}_1 = \emptyset$, generate the masked dataset $\{X_i^{(1)}: i \in \mathcal{I}_2\}$ \;
 \For{$j = 2$  \KwTo  $p+1$}{
    Generate the masked dataset $\{X_i^{(1,j)}: i \in \mathcal{I}_2\}$ \;
    Compute $\Delta_j^i = (1-\beta)L\Big(Y_{i},\mu\big(X_i^{(1)}\big)\Big) - L\Big(Y_{i},\mu\big(X_i^{(1,j)}\big)\Big)$ for all $i \in \mathcal{I}_2$ \;
    $n_j^{+} = \sum_{i\in \mathcal{I}_2} 1_{\{\Delta_j^i > 0\}}$ \;
    \If{$\textsc{BinomTest}(n_j^{+},n_2,1/2,\text{greater}) < \alpha$}{
        $\mathcal{S}_1 = \mathcal{S}_1 \cup \{j\}$ \;
        $\mathcal{C}_1[j] = [\Delta_j^{\lfloor \frac{n_2+1}{2}-q_{1-\alpha/2}\frac{\sqrt{n_2}}{2}\rfloor},\Delta_j^{\lceil \frac{n_2+1}{2}+q_{1-\alpha/2}\frac{\sqrt{n_2}}{2}\rceil}]$ \;
    }
 }
 \caption{First-order SFIT}
 \label{algo 1}
\end{algorithm}

This method is generalized to higher-order interactions between features as explained in more details in \cite{horel2019computationally}.

\section{Simulation Experiments}

\subsection{Data}

\begin{figure}
\includegraphics[width=\columnwidth]{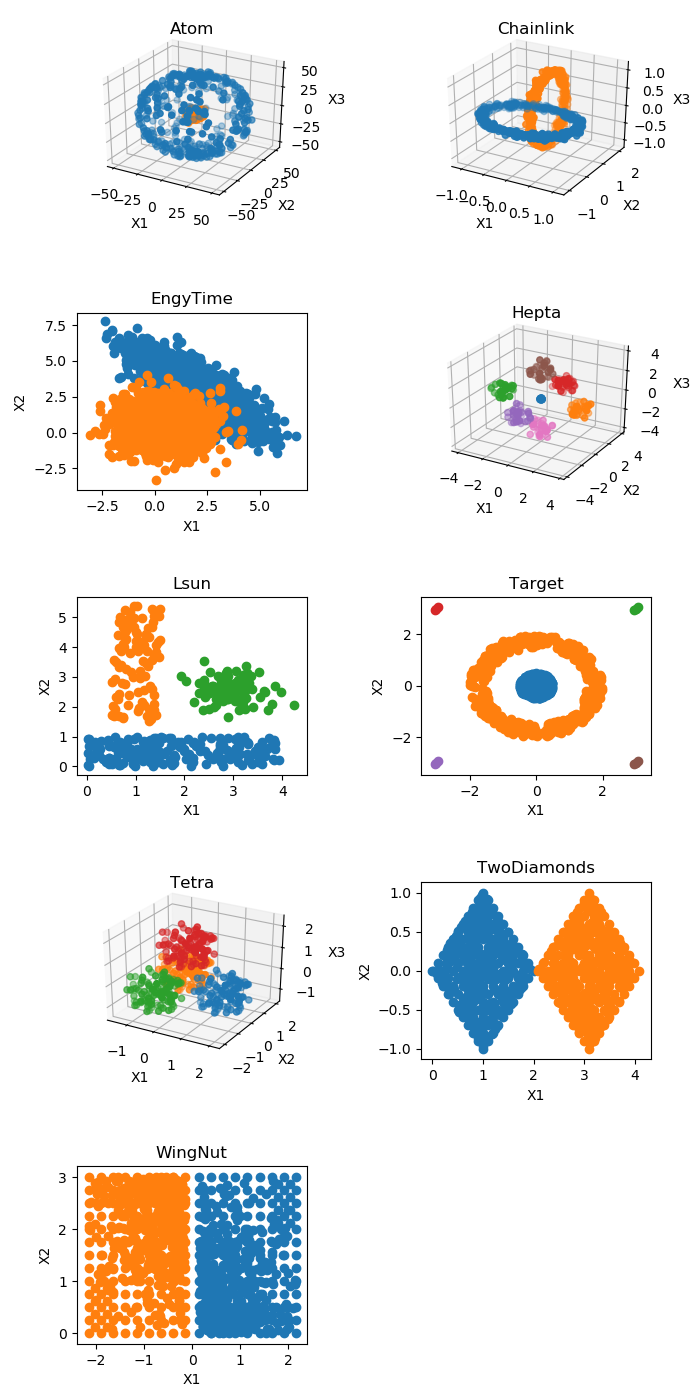}
\caption{Scatter plot representations of the 9 FCPS datasets.}
\label{fig:FCPS_scatterplots}
\end{figure}

In this section, we analyze the performance of our method using the Fundamental Clustering Problems Suite (FCPS) \cite{ultsch2005fundamental}. This suite consists of 9 synthetic datasets commonly used as a benchmark for clustering algorithms. Each dataset consists of either 2 or 3 features along with the cluster labelling for each data point. The datasets can be visualized in Figure \ref{fig:FCPS_scatterplots}.

\subsection{Method}

The purpose of our method is to provide explanations on the clusters returned by a given clustering algorithm. The user can choose any clustering algorithms considered as relevant for the task at hand. The core of our method lies in the explaining step where a classifier trained on the cluster labels is explained through feature importance and significance analysis. In order to assess the performance of our explaining method independently of the chosen clustering algorithm, we decide to use the true cluster labelling provided in the FCPS. 

We train a 3 hidden layers fully connected neural-network to perform classification on the clusters.
This network has ReLU activation functions, a first hidden size of 50, a second hidden size of 25 and a third of 10. The network is trained for at most 50 epochs using Adam optimizer and early stopping. Each dataset is randomly split into a training set used to fit the model and a test set which is used to run the SFIT method.  

%\vspace{-1.3cm}

\subsection{Results}

%\vspace{-1.5cm}

Table \ref{tab:simu_2D} presents the results of the SFIT method applied to the neural networks trained on the 2D datasets of FCPS. For each dataset, we identify which are the significant features and measure the relative importance of single features and interactions. The two datasets TwoDiamonds and WingNut only have the first feature as relevant for classification over the clusters. Our method correctly identifies this pattern. Indeed, only the first feature is returned as significant and the interaction term does not add significant predictive power over the first feature alone in both cases. The three other 2D datasets EngyTime, Lsun and Target all require the combination of the two features to classify the data according to their clusters. As shown in Table \ref{tab:simu_2D}, our procedure indeed consider both features as significant and the predictive importance of the interaction term is always bigger than the importance of single features in all cases. 

\begin{table}
\begin{center}
\caption{Significance and importance of features and interactions of features returned by SFIT method applied on the 2D datasets of FCPS.}
\begin{tabular}{rp{1.5cm}ccc}
\toprule
Dataset   & Significant features & X1 & X2 & (X1, X2) \\ \midrule
EngyTime & 1, 2 & 0.692  & 0.996 & 1.18           \\
Lsun     & 1, 2  & 0.167 & 0.248 & 0.292               \\
Target   & 1,2 & 5e-4  & 3e-4  & 1e-3               \\
TwoDiamonds & 1 & 1.302  & NS  & 1.342               \\
WingNut     & 1 & 1.909  & NS   & 1.899               \\ \bottomrule
\end{tabular}
\label{tab:simu_2D}
\end{center}
\end{table}

Table \ref{tab:simu_3D} presents the results of the SFIT method applied to the neural networks trained on the 3D datasets of FCPS. In all these cases, all features are descriptive and the combination of all three is necessary to identify the different clusters. Our method is able to correctly uncover these patterns since every features are returned as significant in all cases and the interaction effect that involves all three features is always the most important one.

\begin{table}
\begin{center}
\caption{Significance and importance of features and interactions of features returned by SFIT method applied on the 3D datasets of FCPS.}
\setlength\tabcolsep{2.4pt}
\begin{tabular}{rp{1.5cm}ccc}
\toprule
Dataset   & Significant features & X1 & X2 & X3 \\ \midrule
Atom & 1, 2, 3 & 0.002  & 0.002 & 0.002  \\
Chainlink  & 1, 2, 3  & 0.056 & 0.442 & 0.174  \\
Hepta   & 1, 2, 3 & 0.003  & 0.006  & 0.006  \\
Tetra & 1, 2, 3 & 0.311  & 0.483  & 0.676     \\ 
\toprule
Dataset   & (X1, X2) & (X2, X3) & (X1, X3) & (X1, X2, X3) \\ \midrule
Atom & 0.004 & 0.003  & 0.002 &  0.006    \\
Chainlink     & 1.036  & 1.404 & 2.878 & 2.986      \\
Hepta   & 0.027 & 0.067  & 0.0529  &  0.0799  \\
Tetra & 1.511 & 1.299  & 2.167 &  2.574  \\ \bottomrule
\end{tabular}
\label{tab:simu_3D}
\end{center}
\end{table}

Overall our method is able to successfully explain data grouped into clusters. The user can identify what are the features that significantly describe the clusters and hence that can be used to distinguish them from one another. In addition, comparing the importance of single features with the importance of interaction of features sheds light into the structure and geometry of the clusters.

\section{Empirical Application: Financial Ratios Clustering}

\subsection{Data} 
To illustrate our explainable clustering method on a real dataset, we use the Financial Ratios Firm Level dataset from Wharton Research Data Services (WRDS). This dataset provides, for all U.S. companies, 71 commonly used financial ratios grouped into the following seven categories: capitalization, efficiency, financial soundness/solvency, liquidity, profitability and valuation. From this database, we extract a subset of 682 companies which corresponds to all the unique companies listed over the last 10 years. In addition, we have for each company, its NAICS (North American Industry Classification System) and description. The data are centered and scale to unit variance as a pre-processing step.

\subsection{Results}

We cluster our dataset of companies into 5 clusters using an agglomerative hierarchical clustering algorithm. This algorithm works in a bottom-up fashion: each observation starts in its own cluster, and then, clusters are successively merged together. We use a Ward linkage that minimizes the sum of squared differences within all clusters, and euclidean distance. We obtain the following clusters:
\begin{itemize}
\item cluster 1: 201 samples (manufacturing, retails),
\item cluster 2: 277 samples,
\item cluster 3: 139 samples (energy, resources),
\item cluster 4: 60 samples (telecommunication, technology),
\item cluster 5: 5 samples.
\end{itemize}
Because cluster 5 does not have a significant enough size to perform meaningful analysis, we choose to discard it.
By looking at the industry code of the companies of each cluster, we notice that the largest cluster contains a mix of various industries while the 3 remaining clusters are fairly specialized as listed above.

We then label each company using its cluster assignment. We train a 3 hidden layers fully connected neural network to perform classification on the clusters. The dataset is split into three parts, a training set of size 480, a validation set of size 125 and a test set of size 77. We optimize the architecture of the network through random search over the validation set. We end up using ReLU as activation function and a first hidden size of 100, a second hidden size of 50 and a third of 25. The network is trained for at most 50 epochs using Adam optimizer and early stopping. We obtain a classification accuracy of 0.88 on the test set.

We finally run on the trained network one SFIT method per cluster, i.e. by only using the data of this cluster, and returns the five most important features. For the first cluster, the first five features are: gross profit margin, asset turnover, long term debt, current debt and net profit margin, the value of their corresponding test statistic along with their 95\% confidence interval can be found in Table \ref{tab:cluster1}. For the second cluster, they are: book/market, free cash flow/operating cash flow, pre-tax return on total earning assets, sales/invested capital, total debt/ebitda as displayed in Table \ref{tab:cluster2}. For the third cluster: total debt/ebitda, operating profit margin before depreciation, pre-tax return on total earning assets, free cash flow/operating cash flow, return on assets as shown in Table \ref{tab:cluster3}. And for the last cluster: research and development/sales, cash balance/total liabilities, gross profit margin, cash ratio, operating cf/current liabilities as displayed in Table \ref{tab:cluster4}. From these, we can see that the most predictive features are not the same for each cluster. They capture what makes a cluster distinct from the others. This is an efficient way to interpret and explain what are the intrinsic characteristics of a cluster. These results are also consistent with the main types of industries present in each cluster. For instance, the research and development/sales ratio is the most significant feature of the cluster that mainly contains telecommunication and technology companies.

We compare the results obtained from our procedure with the method that consists of representing a cluster by its centroid. Both methods are comparable in the sense that they can both be applied to any clustering algorithms. We define the centroid as the mean of all observations within a cluster. In our case, this returns a vector of dimension 71 that describes the average behavior of the cluster. To make this result more comparable with our, we look for the features of the centroid whose values are the furthest from the mean value over the whole dataset. Formally, for each centroid $c$ and each feature $j$, we compute a score of difference: $$\mathcal{D}_j^c = \frac{|\Bar{X}_j^c - \Bar{X}_j|}{\hat{\sigma}_j}$$ where $\Bar{X}_j$ is the mean value of feature $j$ over the whole dataset, $\Bar{X}_j^c$ is the mean value of feature $j$ over cluster $c$ which is also equal to the value of variable $j$ of the centroid of the cluster and $\hat{\sigma}_j$ is the standard deviation of feature $j$ over the whole dataset. We display in Tables \ref{tab:cluster1_centroid} to \ref{tab:cluster4_centroid} the top 10 features having the largest score of difference $\mathcal{D}_j^c$ for each cluster. In these tables, the features whose name appear in bold are also part of the top 5 most significant variables as identified by our method. Over the 4 clusters, it can be seen that there is a large overlap between these two selected subsets of features with in average 4 out of the 5 most important features selected by SFIT that are also present in the top 10 features as measured by the score of difference. The SFIT has the advantage of providing rigorous significance statistical tests that can be used for feature selection and confidence intervals of the metric that measure feature importance.

\begin{table}
\begin{center}
\caption{Cluster 1 top 5 most significant variables along with their corresponding test statistics and 95\% confidence intervals (CI).}
\begin{tabular}{rccc}
\toprule
Variable   & Median & CI lower bound & CI upper bound \\ \midrule
gpm        & 0.755  & 0.657               & 0.835               \\
at\_turn   & 0.677  & 0.554               & 0.850               \\
lt\_debt   & 0.556  & 0.366               & 0.615               \\
curr\_debt & 0.542  & 0.391               & 0.728               \\
npm        & 0.532  & 0.509               & 0.568               \\ \bottomrule
\end{tabular}
\label{tab:cluster1}
\end{center}
\end{table}

\begin{table}
\begin{center}
\caption{Cluster 2 top 5 most significant variables along with their corresponding test statistics and 95\% confidence intervals (CI).}
\setlength\tabcolsep{4.3pt}
\begin{tabular}{rccc}
\toprule
Variable        & Median & CI lower bound & CI upper bound \\ \midrule
bm              & 0.114  & 0.102               & 0.126               \\
fcf\_ocf        & 0.057  & 0.055               & 0.059               \\
pretret\_earnat & 0.043  & 0.024               & 0.061               \\
sale\_invcap    & 0.037  & 0.035               & 0.040               \\
debt\_ebitda    & 0.031  & 0.021               & 0.039               \\ \bottomrule
\end{tabular}
\label{tab:cluster2}
\end{center}
\end{table}

\begin{table}
\begin{center}
\caption{Cluster 3 top 5 most significant variables along with their corresponding test statistics and 95\% confidence intervals (CI).}
\setlength\tabcolsep{4.3pt}
\begin{tabular}{rccc}
\toprule
Variable        & Median & CI lower bound & CI upper bound \\ \midrule
debt\_ebitda    & 0.856  & 0.746               & 0.987               \\
opmbd           & 0.692  & 0.360               & 0.903               \\
pretret\_earnat & 0.518  & 0.489               & 0.545               \\
fcf\_ocf        & 0.419  & 0.235               & 0.656               \\
roa             & 0.382  & 0.337               & 0.422               \\ \bottomrule
\end{tabular}
\label{tab:cluster3}
\end{center}
\end{table}

\begin{table}
\begin{center}
\caption{Cluster 4 top 5 most significant variables along with their corresponding test statistics and 95\% confidence intervals (CI).}
\begin{tabular}{rccc}
\toprule
Variable    & Median & CI lower bound    & CI upper bound   \\ \midrule
rd\_sale    & 1.89   & 1.70   & 2.02 \\
cash\_lt    & 1.23  & 1.09  & 1.54 \\
gpm         & 0.718  & 0.564   & 0.796 \\
cash\_ratio & 0.449 & 0.333 & 0.527 \\
ocf\_lct    & 0.283  & 0.119 & 0.598 \\ \bottomrule
\end{tabular}
\label{tab:cluster4}
\end{center}
\end{table}

\begin{table}
\begin{center}
\caption{Cluster 1 top 10 features with highest score of difference $\mathcal{D}_j^c$.}
\begin{tabular}{rc}
\toprule
Variable & Score of difference $\mathcal{D}_j^c$  \\\midrule
opmbd                        & 0.830108 \\ 
\textbf{at\_turn   }                  & 0.815331 \\
opmad                        & 0.793995 \\
cfm                          & 0.767334 \\
sale\_invcap                 & 0.725865 \\
ps                           & 0.687671 \\
\textbf{curr\_debt }                  & 0.685457 \\
\textbf{npm }                         & 0.660184 \\
\textbf{gpm }                         & 0.658285 \\
ptpm                         & 0.656759 \\ \bottomrule
\end{tabular}
\label{tab:cluster1_centroid}
\end{center}
\end{table}

\begin{table}
\begin{center}
\caption{Cluster 2 top 10 features with highest score of difference $\mathcal{D}_j^c$.}
\begin{tabular}{rc}
\toprule
Variable & Score of difference $\mathcal{D}_j^c$  \\\midrule
\textbf{bm}              & 0.477207 \\
\textbf{pretret\_earnat} & 0.471108 \\
pretret\_noa    & 0.390561 \\
\textbf{fcf\_ocf}        & 0.309577 \\
ptb             & 0.273166 \\
aftret\_invcapx & 0.268130 \\
lt\_debt        & 0.268081 \\
roa             & 0.264315 \\
lt\_ppent       & 0.227883 \\
roce            & 0.227369 \\ \bottomrule
\end{tabular}
\label{tab:cluster2_centroid}
\end{center}
\end{table}

\begin{table}
\begin{center}
\caption{Cluster 3 top 10 features with highest score of difference $\mathcal{D}_j^c$.}
\begin{tabular}{rc}
\toprule
Variable & Score of difference $\mathcal{D}_j^c$ \\\midrule
curr\_debt      & 0.959909 \\
\textbf{fcf\_ocf}      & 0.950367 \\
\textbf{debt\_ebitda}    & 0.900283 \\
GProf           & 0.855419 \\
at\_turn        & 0.738199 \\
\textbf{opmbd}           & 0.675834 \\
bm              & 0.674283 \\
\textbf{pretret\_earnat} & 0.672521 \\
cfm             & 0.652589 \\
\textbf{roa }            & 0.630866 \\ \bottomrule
\end{tabular}
\label{tab:cluster3_centroid}
\end{center}
\end{table}

\begin{table}
\begin{center}
\caption{Cluster 4 top 10 features with highest score of difference $\mathcal{D}_j^c$.}
\begin{tabular}{rc}
\toprule
Variable & Score of difference $\mathcal{D}_j^c$   \\\midrule
\textbf{cash\_lt}     & 2.143874 \\
\textbf{rd\_sale}     & 2.126812 \\
\textbf{cash\_ratio}  & 2.011226 \\
quick\_ratio & 1.714341 \\
\textbf{gpm}          & 1.399909 \\
curr\_ratio  & 1.329093 \\
ps           & 1.254171 \\
cash\_debt   & 1.102644 \\
ptpm         & 1.085758 \\
debt\_assets & 1.085222 \\ \bottomrule
\end{tabular}
\label{tab:cluster4_centroid}
\end{center}
\end{table}

We propose in this paper a novel two-steps method that can interpret any clustering algorithms. For each cluster, this method identifies the statistically significant features that characterize it as well as feature interactions. We justify the necessity of such a method by describing a Wealth Management Compliance use-case that requires explaining clusters of under-performing accounts. We demonstrate its effectiveness on simulated data and a real dataset of financial ratios of U.S. companies.

\begin{acks}
The authors would like to thank the Wealth Management Technology team at J.P. Morgan and especially Amish Seth for their collaboration and help to gain insight on the business case.  
\end{acks}

\bibliographystyle{ACM-Reference-Format}
\bibliography{biblio}

\end{document}